\documentclass[conference]{IEEEtran}
\IEEEoverridecommandlockouts

\usepackage{color,soul}
\usepackage{optidef}
\usepackage[ruled,vlined]{algorithm2e}
\usepackage{cite}
\usepackage{subfig}
\usepackage{svg}
\usepackage{amsmath,amssymb,amsfonts}
\usepackage{algorithmic}
\usepackage{graphicx}
\usepackage{textcomp}
\usepackage{xcolor}
\usepackage{textcomp}
\usepackage{verbatim}
\usepackage{multirow}
\usepackage[normalem]{ulem}
\useunder{\uline}{\ul}{}

\def\BibTeX{{\rm B\kern-.05em{\sc i\kern-.025em b}\kern-.08em
    T\kern-.1667em\lower.7ex\hbox{E}\kern-.125emX}}

\begin{document}

\title{A Learning Framework for Bandwidth-Efficient Distributed Inference in Wireless IoT}

\author{
\IEEEauthorblockN{Mostafa Hussien\textsuperscript{1,2}, Kim Khoa Nguyen\textsuperscript{2}, and Mohamed Cheriet\textsuperscript{2}}
\IEEEauthorblockA{
\textsuperscript{1} Resilient Machine-learning Institute (ReMI)\\
\textsuperscript{2} École de technologie supérieure (ÉTS), Univeristy of Québec\\ Montréal, QC, Canada}
}

\maketitle

\begin{abstract}
In wireless Internet of things (IoT), the sensors usually have limited bandwidth and power resources. Therefore, in a distributed setup, each sensor should compress and quantize the sensed observations before transmitting them to a fusion center (FC) where a global decision is inferred. Most of the existing compression techniques and entropy quantizers consider only the reconstruction fidelity as a metric, which means they decouple the compression from the sensing goal. In this work, we argue that data compression mechanisms and entropy quantizers should be co-designed with the sensing goal, specifically for machine-consumed data. To this end, we propose a novel deep learning-based framework for compressing and quantizing the observations of correlated sensors. Instead of maximizing the reconstruction fidelity, our objective is to compress the sensor observations in a way that maximizes the accuracy of the inferred decision (i.e., sensing goal) at the FC. Unlike prior work, we do not impose any assumptions about the observations distribution which emphasizes the wide applicability of our framework. We also propose a novel loss function that keeps the model focused on learning complementary features at each sensor. The results show the superior performance of our framework compared to other benchmark models.
\end{abstract}


\section{Introduction}
\label{sec:intro}

\begin{figure}
    \centering
    \includegraphics[width=0.45\textwidth]{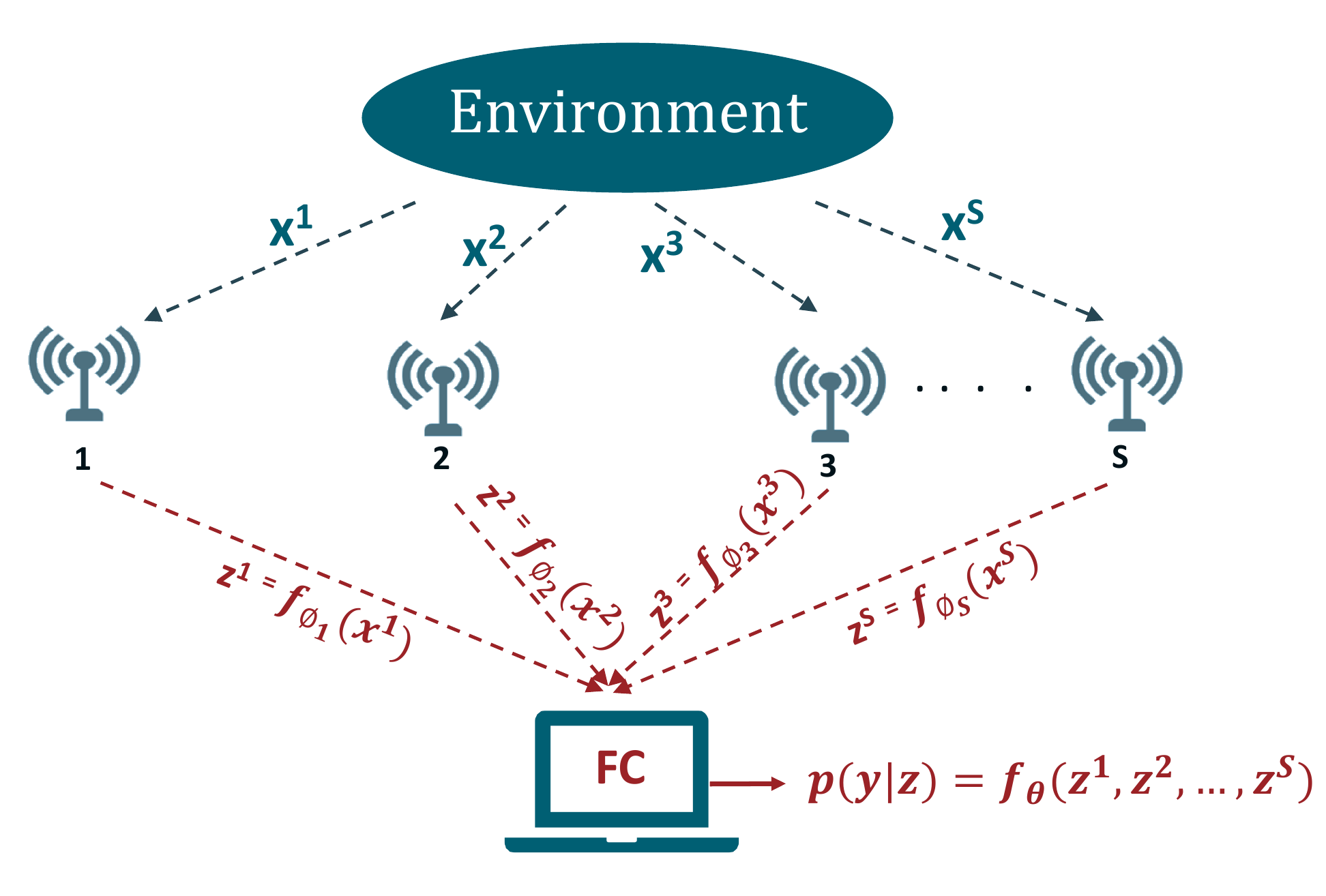}
    \caption{Multiple correlated sensors transmitting compressed quantized form, z, of the sensed data to a FC which applies a decision rule based the aggregated data.}
    \label{fig:WSN_figure}
\end{figure}

\IEEEPARstart{M}{any} wireless Internet of things (IoT) applications employ a distributed inference mechanism e.g., radar systems, surveillance video, or multi-sensory human activity recognition. In the later system for example, a human wears multiple, spatially-distributed, sensors (e.g., gyroscope and accelerometer). A decision about the human activity (e.g., walking, running, etc.) is inferred from the received sensor signals. Fig. \ref{fig:WSN_figure} shows this general scenario. In such a scenario, if each sensor considered only its local observations for inferring decision, the error probability would be much higher compared to the scenario in which a global decision is taken based on aggregated sensor data \cite{salehkalaibar2018hypothesis}.

To tackle this problem, a distributed setup may be employed in which the sensed data (a.k.a, environment observations) are sent to a central node, called \textbf{fusion center (FC)}. The FC infer a global decision based on the data aggregated from all sensors. However, the sensors usually have limited power and bandwidth resources. For example, each sensor may assigned a fixed data rate of, $R$ bps. Therefore, each sensor should compress and quantize its observation to fit the assigned bit rate before transmission. The FC, then, performs a specific inference task (i.e., the sensing goal). Given that the compression and quantization of raw observations introduce some information loss, the FC takes the decision based on partial information. This may result in a reduced decision accuracy at the FC \cite{salehkalaibar2018hypothesis}. Optimally processing the observations at each sensor can minimize the loss in the decision accuracy \cite{abdi2020max}. For conditionally-independent sensor observations, an optimal decision can be easily reached using Bayesian inference theory \cite{zhu2019cost}. However, the conditional-independence assumption does not hold for many real-life problems. In prior work \cite{chamberland2007wireless, tay2009bayesian}, the observations distribution is assumed to be known at the FC. In this case, the goal is to design an optimal decision rule that maximizes the likelihood of the correct decisions. Unfortunately, in many practical applications, this distribution is not known in advance which increases the problem complexity. In this case, data-driven solutions can be introduced as robust and feasible alternatives.

Although different works in the literature propose compression and quantization techniques for sensor data, their goal was mainly obtaining a high-fidelity reconstruction at the FC \cite{hussien2021fault}. This seems relevant for human-consumed data such as images and videos. However, for machine-consumed data, adopting the reconstruction fidelity as a metric is doubtful. Indeed, the accuracy of the automated decisions taken based on the received data is the most crucial factor.

In this work, we tackle the problem of compressing and quantizing correlated-sensor observations for distributed inference tasks. Our main objective is maximizing the accuracy of the inferred decision rather than minimizing the reconstruction loss. While most of the literature work assume sensors independence for mathematical tractability, we address the more challenging scenario of correlated sensors. We argue that this correlation can be exploited to obtain higher compression ratios without considerable loss in the decision accuracy. These higher compression ratios can be obtained by transmitting the unique features of each sensor, and avoid transmitting redundant features which are likely be transmitted by other nodes in the network. In other words, we can formulate our research question as: can we distributively screen redundancies in sensor observations to transmit only informative data without imposing any assumptions on the observations distribution?

To answer this question, we exploit the current revolution in statistical learning techniques, especially deep learning. We propose a novel deep learning framework for compressing and quantizing the observations at each sensor. Ina addition, the framework is jointly trained with the decision rule at the FC in an end-to-end fashion to maximize the accuracy of the inferred decision. End-to-end learning refers to training a possibly complex learning system by applying gradient-based learning to the system as a whole \cite{glasmachers2017limits}. We also propose a new loss function for the proposed framework that helps the sensors learn decision-aware representation for the observations. Furthermore, we propose a training algorithm to efficiently train the proposed framework. Extensive results show the robustness and superiority of our proposed framework compared with different benchmark models. 

\paragraph{Contribution}
This paper presents a novel deep learning-based compression framework for correlated-sensors data. Discrete representation's autoencoders are adopted at each sensor to generate the compressed quantized form of the observations. At the FC, a multi-layer perceptron (MLP) architecture is adopted to jointly learn the decision rule with the sensor encoders. The main contribution of this work comes in three folds:
\begin{enumerate}
    \item Extending autoencoders to learn a compressed and quantized representation for correlated-sensor observations. This learned representation conveys the complementary features at each sensor observation which help maximizing the likelihood of the correct decision at the FC while satisfying a communication constraint. This representation is jointly learned with the decision-rule at the FC in an end-to-end fashion to maximize the decision accuracy. 
    \item Proposing a novel loss function that encourages the model to learn the unique features at each sensor. The function learns the soft probabilities of a baseline model trained to predict $P(Y|X)$. Moreover, we present a training algorithm that efficiently works in a wide range of applications.
    \item Eliminating the conditional-independence assumption between sensor observations which has been widely adopted for mathematical tractability. Beyond most of the literature work which, for the sake of simplicity, assume only binary hypothesis problems, we consider the more complicated scenario of multi-hypothesis problems.
\end{enumerate}

The rest of this paper is organized as follows: Section \ref{sec:problem_statement} formulates the problem. In Section \ref{sec:proposed_framework}, we describe the various elements of the proposed framework. The discussion and the experimental results are given in Section \ref{sec:results_and_discussion}. Section \ref{sec:related_work} briefly summarizes the relevant work in the literature, while Section \ref{sec:conslusion} concludes our work. 

\section{Problem Statement}
\label{sec:problem_statement}

\noindent \textbf{Notation:} Through this text, we refer to random variables by capital letters (e.g. $X$). Small letters refer to one realization of a random variable (e.g., $x$). Superscripts denote the sensor number. For example, $x^i$ denotes the observation at sensor $i$. The observations are referred to by $X$ while $Y$ refers to the random variable of the labels (i.e., the target decisions at the FC). The parameters of the encoder at the $i^{th}$ sensor is referred to as $\phi_i$. The parameters of the decision rule at the FC is referred to by $\omega$. The $\log(\cdot)$ function uses a base of 2. Table \ref{Table:notation} summarizes the used symbols and notations.

\begin{table}[t]
\caption{The notation used through the text.}
\resizebox{\columnwidth}{!}{
\normalsize
\begin{tabular}{p{0.20\linewidth} | p{0.8\linewidth}}
\hline \hline
\multicolumn{1}{p{0.20\linewidth}|}{\centering \textbf{Symbols}} & \multicolumn{1}{p{0.8\linewidth}}{\centering \textbf{Description}}                                           \\ \hline \hline
\centering $x^i$                                              & The current observation captured by sensor $i$.                                        \\ \hline
\centering $z^i$                                              & The compressed and quantized representation for the current observation at sensor $i$. \\ \hline
\centering $y$                                                 & The true label (or decision) to be predicted at the FC.                  \\ \hline
\centering $\hat{y}$                        & The predicted label at the FC.                                                             \\ \hline
\centering $\phi_i$                         & The encoder parameters at sensor $i$.                                                      \\ \hline
\centering $\theta$                         & The parameters of the decision function (i.e., decision rule) at the FC.                   \\ \hline
\centering $\omega$                         & The parameters of the decision function given the raw-observations.                       \\ \hline
\centering $S$                              & The total number of sensors.               
                                            \\ \hline
\centering $d$                              & The dimension of the raw observations.
                                            \\ \hline 
\centering $n$                              & The dimension of the compressed and quantized observations.
                                            \\ \hline
\centering $R$                              & The bandwidth (in bps) assigned to each sensor.
                                            \\ \hline
\centering $\chi$                           & The observation space, $\mathbb{R}^d$. 
                                            \\ \hline 
\centering $Z$                              & The latent space, $\{0,1\}^n$.
                                            \\ \hline
\centering $f_{\phi_i}$                         & The encoder  function at the $i^{th}$ sensor given by a neural network  parameterized by parameters $\phi_i$.                          \\ \hline
\centering $g_{\theta}$                       & The decision rule at the FC given by a neural network parameterized by parameters $\theta$.                          \\ \hline
\centering $\mathbb{S}^n$                   & An n-dimensional vector where each element belongs to the set $\mathbb{S}$.            \\ \hline \hline
\end{tabular}
}
\label{Table:notation}
\end{table}

Suppose $Y$ is a discrete random variable, representing a hypothesis about an environment. The variable takes values in: $y \in \left \{ 1,2, \dots, C \right \}$ where $C$ is the number of possible hypothesis or classes. Our goal is to form an estimate, $\hat{Y}$, of the true hypothesis, based on a set of observations collected from a set of $S$ sensors. Accordingly, for each $t=1,\dots,S$, let $x^t$ represents the observation at node $t$, where $x^t \in \mathbb{R}^d$ in some space $\chi$ known as the observation space. The set of all observations correspond to an $S$-dimensional random vector $X=(x^1,x^2, \dotsc ,x^S) \in \chi^S$ drawn from the conditional distribution $P(X|Y)$. 

Our objective is to reach an optimal estimate $\hat{Y}$ for the true labels $Y$ at the FC. If the FC has access to the observations distribution, $P(X|Y)$, then an optimal decision rule can be easily formulated. For example, with binary hypothesis, an optimal decision rule can be reached by means of likelihood ratio test: $P(X|Y=1)/P(X|Y=-1)$. However, in real-world problems, the FC does not know the distribution of the observation a priori, and it has access to only summarized forms of the original observations, $z^t$, for all values of $t$. More specifically, we assume that each sensor, $t$, is restricted to a given bandwidth of, $R$, bps. Therefore, each sensor is allowed to transmit an \textit{n-dimensional} message, $z^t \in \{0,1\}^n$, taking values in some space $Z$, such that $n \leq R$. The conversion from the observation space, $\chi$,  to $Z$-space is carried out by an encoder $q: \chi \rightarrow Z$. The encoder, $q$, maps an input observation, x, in $\chi$-space, to a codeword, $z$, in $Z$-space. This encoded observation, $z$, will be sent to the FC. To compute the estimate $\hat{Y}$, the FC applies a certain decision rule, $g_{\theta}$, on the aggregated received messages such that $\hat{Y}= g_{\theta} ( z^1,z^2, \dots ,z^S)$. It is known from the rate-distortion theory that the rate, $R$, and the distortion at the receiver (in terms of reconstruction loss) are inversely proportional. Therefore, larger rate, $R$, implies better reconstruction fidelity at the receiver end. However, in our problem we are not concerned about the reconstruction fidelity as our main objective. Rather, we are more interested in maximizing the accuracy of the inferred decisions. 


Inherently, increasing the rate, $R$, will increase the information included in a message, $z^t$, which increases the FC accuracy. In other words, increasing the rate, $R$, increases the mutual information, $I$, between the joint distributions $P(\hat{Y}|Z)$ and $P(\hat{Y}|X)$. However, for limited bandwidth systems, increasing the bandwidth is not an available option and each sensor should commit to the assigned bandwidth of $R$ bps. In this case, for correlated sensor observations, the redundancy between the different sensor observations can be exploited to obtain more efficient compression with minimal loss in the decision accuracy at the FC. This can be expressed by the optimization function given in (\ref{eq:opt_func1}).

\begin{equation}
\begin{aligned}
\overset{min}{\phi , \theta}\quad & \frac{1}{N} \sum_{j=1}^{N}-\log (g_\theta(z_j)=y_j)\\
\textrm{s.t.} \quad & z_j=(f_{\phi_1} (x_1), f_{\phi_2}(x_2), \dots, f_{\phi_S}(x_S)),\\
&f_{\phi_i} \in \{ 0,1\}^n \;\; \forall i \in \{1,2,\dots,S\},\\
&n \leq R
\end{aligned}
\label{eq:opt_func1}
\end{equation}

\noindent where $N$ is the total number of points in a test set, $g_\theta$ is the decision function at the FC parameterized by parameters $\theta$, $f_{\phi_i}$ is the encoder function at the $i^{th}$ sensor parameterized by $\phi_i$, and $R$ is the bandwidth assigned for each sensor. The same objective can be formulated in terms of the \textit{Kullback–Leibler} divergence between the two conditional distributions of the decision given the raw observations and the compressed messages as given in (\ref{eq:opt_func2}) and (\ref{eq:opt_func3}).

\begin{equation}
\small
    \begin{aligned}
    \overset{\min}{\omega, \theta, \phi_i} \quad & KL \left ( P(\hat{Y}|X) \left |  \right | P(\hat{Y}|Z) \right ) \\
    \textrm{s.t.} \quad & P( \hat{Y}|X) = f_\omega \left (x_1, x_2, \dots, x_S \right), \\
    & P( \hat{Y}|Z) = g_\theta \left(  f_{\phi_1} (x_1), f_{\phi_2}(x_2), \dots, f_{\phi_S}(x_S) \right), \\
    &f_{\phi_i} \in \{ 0,1\}^n \;\; \forall i \in \{1,2,\dots,S\},\\
    &n \leq R
    \end{aligned}
    \label{eq:opt_func2}
\end{equation}

\noindent where $\omega$ is the parameters of a benchmark model (larger neural network model that trained to classify the original observations without compression). But the \textit{KL-divergence} is given by:

\begin{equation}
\small
KL(P||Q) = \sum_{i}^{} P(i) \log (\frac{P(i)}{Q(i)})
\label{eq:KL_eq}    
\end{equation}

Substituting the $KL$ term in (\ref{eq:opt_func2}) by (\ref{eq:KL_eq}), we get Eq. (\ref{eq:opt_func3}).

\begin{equation}
\small
    \begin{aligned}
    \overset{\min}{\omega, \theta, \phi_i} \quad & \sum_{i}^{} P(\hat{Y}_i|X_i) \log (\frac{P(\hat{Y}_i|X_i)}{P(\hat{Y}_i|Z_i)}) \\
    \textrm{s.t.} \quad & P( \hat{Y}|X) = f_\omega \left (x_1, x_2, \dots, x_S \right), \\
    & P( \hat{Y}|Z) = g_\theta \left(  f_{\phi_1} (x_1), f_{\phi_2}(x_2), \dots, f_{\phi_S}(x_S) \right), \\
    &f_{\phi_i} \in \{ 0,1\}^n \;\; \forall i \in \{1,2,\dots,S\},\\
    &n \leq R
    \end{aligned}
    \label{eq:opt_func3}
\end{equation}

Two main points should be considered. Firstly, the message space $\left \{ 0,1 \right \}$ is significantly smaller than the observation space $\mathbb{R}$. Secondly, the required dimension for the message, $n$, is substantially smaller than that of the raw observation, $d$ (i.e., $n \ll d$). Therefore, the problem can be thought of as finding, for each sensor, $t$, an optimal encoder and quntizer $q$: $q(x^t) = z^t$ that maximizes the mutual information between the two distributions $P(Y|X)$ and $P(\hat{Y}|Z)$ under a certain rate $R$.


\section{Proposed Framework}
\label{sec:proposed_framework}

\begin{figure}[t]
\centerline{\includegraphics[width=0.45\textwidth]{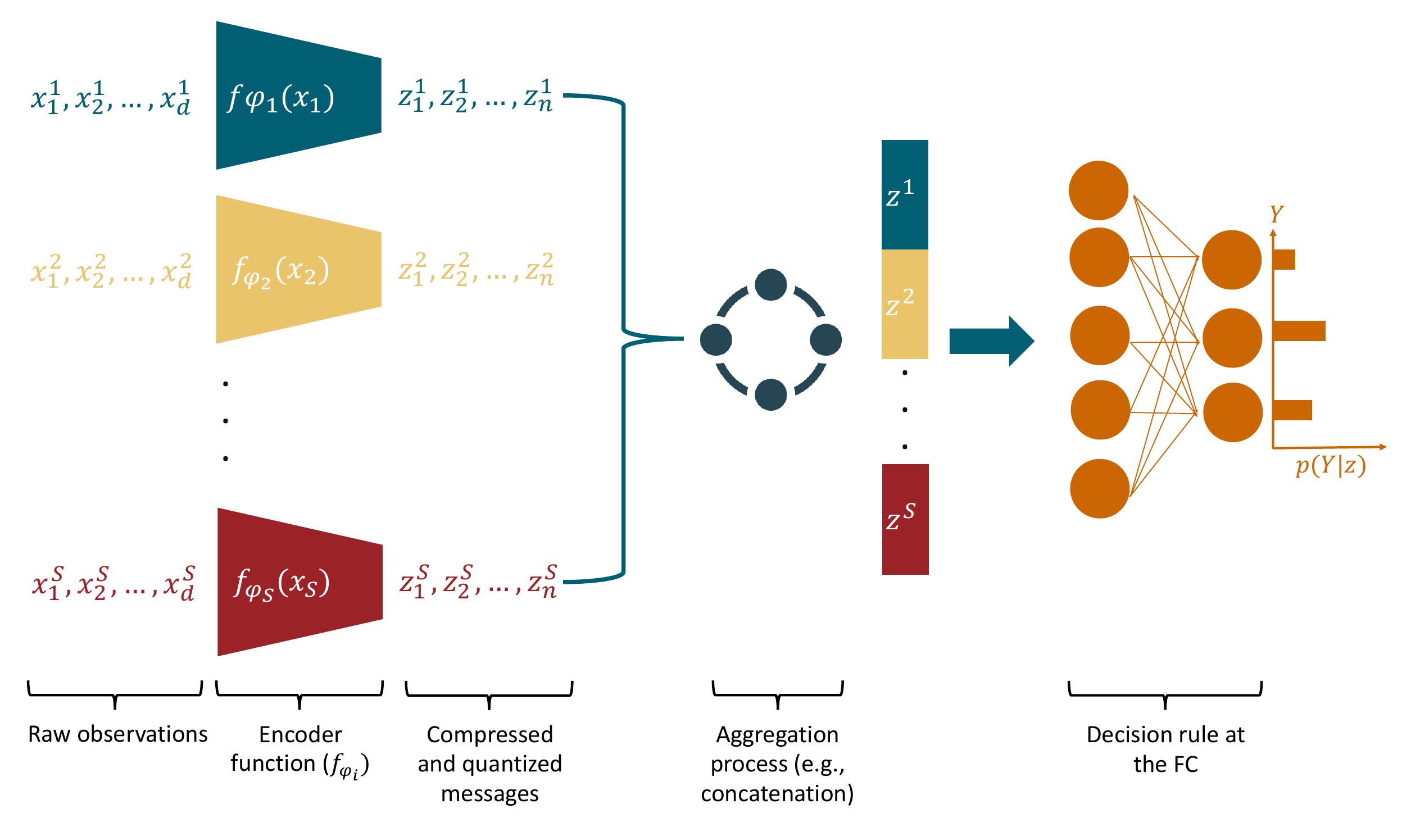}}
\caption{Diagram demonstrating the system model. In the left, we see the sensor observations going through the discrete encoders to obtain the compressed quantized form of the observations. Then these messages are sent to the FC which passes the aggregated message to the neural network architecture to get a hypothesis estimation.}
\label{fig:sys_model}
\end{figure}

\subsection{Autoencoders}

\begin{figure*}[t]
\centerline{\includegraphics[width=0.95\textwidth]{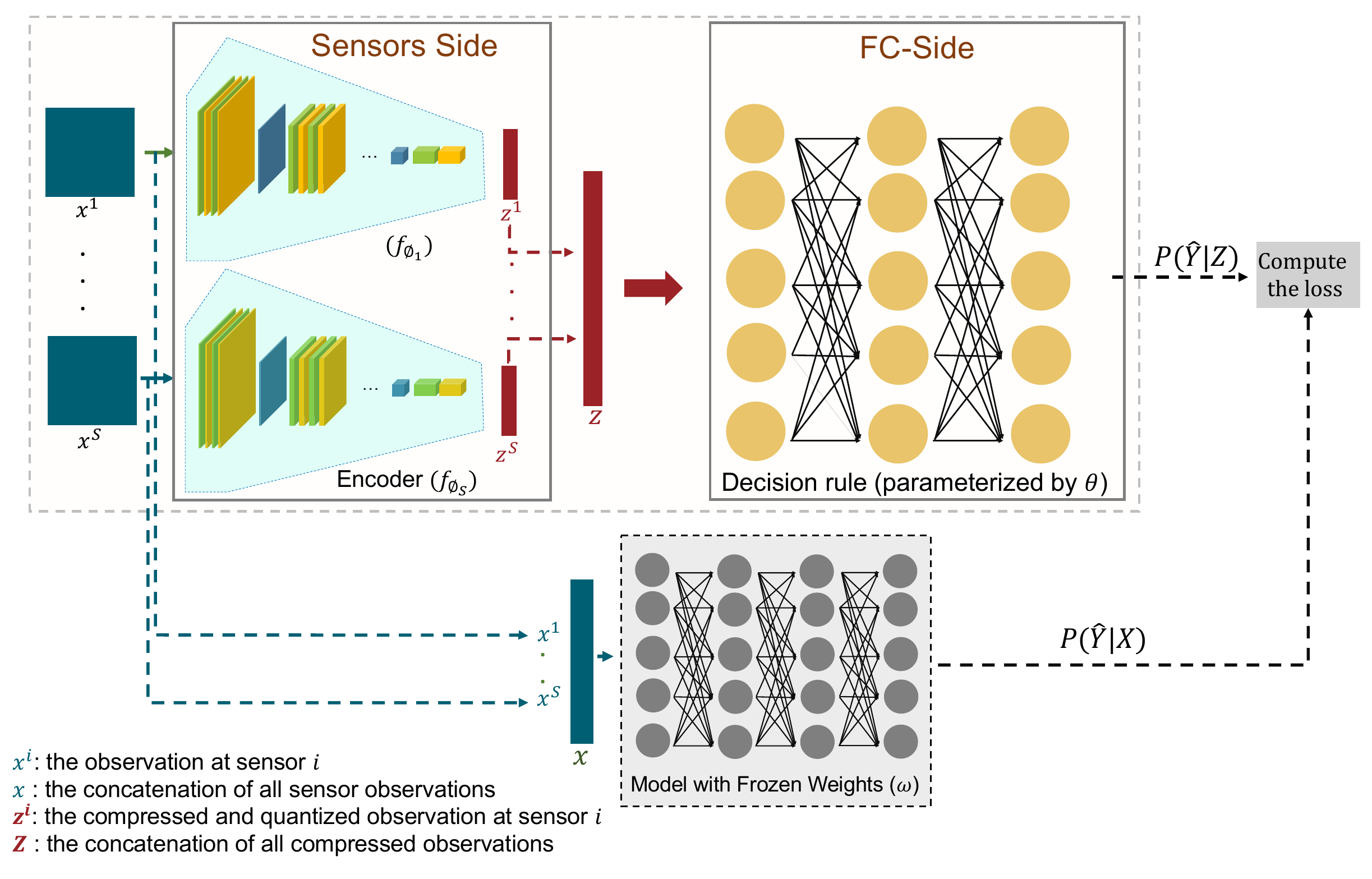}}
\caption{The proposed framework for deep distributed inference in wireless sensor networks.}
\label{fig:system_model}
\end{figure*}

One of the powerful deep-learning architectures that achieved state-of-the-art results in different contexts is autoencoder (AE). AE is a neural-network architecture consisting of two models namely, encoder and decoder. The encoder model is sometimes referred to as the inference model while the decoder model is referred to as the generative model. The encoder maps an \textit{I-dimensional} input to an \textit{O-dimensional} codeword, where $O \ll I$. The decoder then reconstructs the input from this, compressed, codeword. This codeword is usually referred to as \textbf{latent representation} and it belongs to a space called \textbf{latent space}. This process is performed in an end-to-end fashion which implies that the encoder learns to compress the data in a way that help the decoder in the reconstruction process. If the codeword is quantized (binary or multi-level), then the architecture is referred to as \textbf{discrete representation's autoencoder}. For further details on autoencoder architecture, we refer to \cite{majumdar2018blind}.

According to the aforementioned problem formulation, our objective is to jointly learn an optimal encoder and quantizer at each sensor, $q^t$: $q^t(x_i) = z_i$, and an optimal decision rule at the FC $g_{\theta} ( z^1,z^2, \dots, z^S)$. To this end, we adopt a discrete-representation autoencoder at each sensor node to compress and quantize sensor observations, see Fig. \ref{fig:sys_model}. Its worth to differentiate between compression and quantization in this context. By compression, we mean the mapping from a higher-dimensional to a lower-dimensional space, $f: \mathbb{S}^{d}\rightarrow \mathbb{S}^{n}$, where $n \ll d$ and $\mathbb{S}$ is a certain set. On the other hand, quantization is mapping the values of individual dimensions from a set $\mathbb{S}_1$ to a set $\mathbb{S}_2$, where the cardinality of $\mathbb{S}_1$ is smaller than that of $\mathbb{S}_2$, (i.e., $|\mathbb{S}_1| < |\mathbb{S}_2|$). 

Each sensor transmits the output of its \textit{encoder model} to the FC. The output of the encoder model at sensor $i$ is given by: $f_{\phi_i}(\cdot)$. At the FC, an MLP neural network parameterized by parameters, $\theta$, is used to approximate the optimal decision rule, see Fig. \ref{fig:system_model}. The decision rule at the FC is given by:

\begin{equation}
    g_{\theta} ([f_{\phi_1}(x^1), f_{\phi_2}(x^2), \dots, f_{\phi_S}(x^S)]).
\end{equation}

\noindent where $x^i$ is the current observation at the $i^{th}$ sensor.
\subsection{Implementation Details}
The encoder architecture at each sensor is an MLP of three fully-connected layers with \textit{ReLU} activations.  
In the output layer of the encoder, a \textit{QSigmoid} activation is used \cite{moons2017minimum}. 
In the FC, we used six fully connected layers with \textit{ReLU} activations in the hidden layers and \textit{Softmax} activation 
in the output layer. 
The model weights are initialized using \textit{He} initializer \cite{he2015delving}. The models are trained using \textit{Adam} optimizer \cite{kingma2014adam}, with  (0.01) learning-rate and optimized to minimize our proposed loss function given in Eq. (\ref{eq:proposed_loss}). Due to the adopted end-to-end training, the encoders will learn to encode the unique information at each sensor that help the FC to infer the correct decisions. Furthermore, the FC model optimizes its weights to maximize the likelihood of the correct decision given the encoded observations. So, we can interpret the optimization of the classifier weights at the FC as learning an optimized threshold function for the decision rule. 

\subsection{Training Procedure}

\begin{algorithm}[t]
\SetAlgoLined
\KwIn{Dataset $D$, consisting of $N$ tuples of observations acquisted from $S$ sensors and the corresponding label $y$.}
\KwOut{Model parameters, $\theta$, and $\phi_i$ for $i \in \{1,2,\dots,S\}$}

\begin{itemize}
    \item [-] At each sensor, $s_i$, train an autoencoder to reconstruct its input using observations in D;\\
    \item [-] Train an inference model, $I_1$, to approximate the conditional distribution $p(\hat{Y}|X)$;\\
    \item [-] Freeze the weights of $I_1$, known as $\omega$;\\
    \item [-] Train an inference model, $I_2$, (jointly with the encoders weights, $\phi_i$ for $i \in \{1,2,\dots,S\}$) to approximate the conditional distribution $p(\hat{Y}|Z)$;\\
    \item [-] Return the learned parameters of $I_2$ (i.e., $\omega$), along with $\phi_i$ for $i \in \{1,2,\dots,S\}$.
\end{itemize}

\caption{The training procedure for the proposed framework, $S$, sensors.}

\label{algo:training}
\end{algorithm}

The training of the proposed framework comes in three phases. In the first phase, we train an autoencoder at each sensor. The autoencoders are trained for input reconstruction from a compressed codewords by minimizing l2-norm function given in Eq. (\ref{eq:MSE}).

\begin{equation}
    \mathcal{L} = \frac{1}{N} \sum_{i=1}^{N} {\left \| x_i - \hat{x}_i \right \|}^{2}
\label{eq:MSE}    
\end{equation}

In the second phase, we, independently, train an inference model, $I_1$, that takes as input the raw observations, X = [$x_1, x_2, \dots, x_S$], and outputs the corresponding decision. Note that the inputs to this model are the raw observations without compression or quantization. The model is trained to optimize the classical Crossentropy function (\ref{eq:cross_entropy}). 
\vspace{-0.2cm}
\begin{equation}
\label{eq:cross_entropy}
    f(y, \hat{y}) = -\sum_{i=1}^{C} y_i \log (\hat{y}_i) + (1-y_i) \log(1-\hat{y}_i),
\vspace{-0.1cm}
\end{equation}
\noindent where $y_i$ is the target label vector of the $i^{th}$ data point and $C$ is the number of classes. The output of the model $I_1$ approximates the conditional distribution $P(\hat{Y}|X)$. The set of parameters in the model $I_1$, denoted by $\omega$, is then frozen and its will be used only for computing the loss value of the inference model at the FC. This model represent the baseline model that we aim to approximate after the compression and the quantization take place. We elaborate more on this part in subsection \ref{subsec:proposed_loss}.

In the third and last phase, we use the encoder model (of the autoencoders trained in the first phase) at each sensor to compress the observation captured by this sensor. The output of the encoder at sensor $i$ is denoted by $z_i$. The parameters of an encoder model of the autoencoder at sensor, $i$, denoted by $\phi_i$. Accordingly, $z^i=f_{\phi_i}(x)$. The output of all the encoders are concatenated and fed to an inference model, $I_2$, with parameters $\theta$ to predict the output. In this case, the output of $I_2$ approximates the conditional distribution $P(\hat{Y}|Z)$. Its worth noting that the weights of $I_2$, $\theta$, are trained jointly with the encoders weights, $\phi_i$, at each sensor. This means that the training in the last phase is done in an end-to-end fashion between the encoders weights and the parameters of decision rule at the FC. Algorithm \ref{algo:training} summarizes the training procedure. 

\begin{algorithm}[t]
\SetAlgoLined
 Randomly select two random points $x_1$, $x_2$;\\
 Encode each data point using the encoder function, $f_\phi$:\\
 $z_1, z_2 = f_\phi(x_1), f_\phi(x_2)$;\\

 $h = z_1 \oplus z_2$;\\
 i=0;\\
 \While{$i < len(h)$}{
 \If{h[i] equals 1}{
     Flip the bit at $z_1$[i];\\
     i = i + 1;
 }
 $\bar{x_{1i}}= f_\theta(z_1)$; where $f_\theta$ is the decoder function.\\
  Plot $\bar{x_{1i}}$;\\
 }

 \caption{The procedure for evaluating the semantics of the latent codewords.}
 \label{algo:semantics}
\end{algorithm}

\subsection{Proposed Loss Function}
\label{subsec:proposed_loss}

In the first and second phases of the training, we optimize the MSE and Crossentropy loss functions, respectively. However, in the third phase of the training in which we jointly train the sensors encoders and the inference-model at the FC, minimizing the traditional Crossentropy is found to be limited in solving the problem being addressed. Recall from previous section that the objective of the proposed framework is to make the encoders benefit from the redundancies (between the sensor observations) to obtain high compression ratios without harming the decision accuracy (i.e., the sensing goal). This implies that encoders should learn to encode the complementary features of its observation. To this end, we propose a novel loss function (\ref{eq:proposed_loss}).

\begin{equation}
\label{eq:proposed_loss}
    \mathcal{L}(Y,\hat{Y}) = CE(Y,\hat{Y}) + KL (P(\hat{Y}|X)||P(\hat{Y}|Z))
\end{equation}

The proposed function helps the model to learn a joint conditional distribution for the decision given the compressed observations, $P(\hat{Y}|Z)$, which is as similar as possible to the joint conditional distribution for the label given the uncompressed observations, $P(\hat{Y}|X$). This term reduces the loss in the decision accuracy due to the compression of the sensor observations. Given a limited budget of data bits to encode the observations, we argue the proposed function encourages the encoders to encode only the relevant features that help in maximizing the likelihood of the correct decision at the FC.

Note that we handcrafted a model for each dataset to achieve the highest possible accuracy. The models have been selected according to the proposed loss function, Eq. (\ref{eq:proposed_loss}), such that it emphasizes on learning complementary features at each sensor. To this end, the second term in Eq. (\ref{eq:proposed_loss}) adds a regularization term based on the KL-divergence between the conditional probability distribution of the decision given the full observation $P(\hat{Y}|X))$, and the conditional probability distribution of the decision given the compressed and quantized version of the observations $P(\hat{Y}|Z))$. Moreover, our model jointly learns a quantizer function (entropy encoder) along with the source encoder. Jointly learning the encoders with the decision rule encourages the model to learn only the complementary features at each sensor. The proposed models work well with each problem without overwhelming the framework with complex architectures such as \textit{AlexNet}, \textit{ResNet}, \textit{GoogleNet}, etc \cite{lecun2015deep}. The power of these complex models is required mainly for high-dimensional observation space, such as surveillance cameras. In this case, the hidden (deep) convolutional layers can extract spatial features in the observations in an efficient way. However, in the problems with lower-dimensional observation space such as our case, handcrafted models will outperform these models in terms of accuracy and complexity. This conclusion is compatible with the findings reported in \cite{suto2019efficiency}.
\subsection{Dataset Preparation}
The proposed framework is general and widely applicable in different problems. For the framework to be employed in a certain distributed inference task, a dataset should be prepared for training.  A typical dataset consists of $N$ data points along with the associated labels $\{x_i,y_i\}_{i=1}^N$. Each data point, $x_i$, represents the concatenation of simultaneous readings from $S$ sensors such that $x_i = [x^1_i, x^2_i, \dots, x^S_i]$. The label $y_i \in \{1, 2, \dots, C\}$ is the one-hot encoded vector of the target hypothesis (i.e., class) associated with these sensor readings. It is worth to note that these readings are assumed to be perfectly synchronized and each data point represents the readings at the same time step.

\section{Results and Discussion}
\label{sec:results_and_discussion}

We show the results of testing the proposed framework using different datasets. Each dataset represents different environment setting and generating distribution. This section is organized to three subsections. The accuracy of distributed detection problem is presented in subsection \ref{subsec:res_dist_detec}. The semantics of the latent representations is presented in subsection \ref{subsec:res_sem_rep}. The last subsection explores the efficiency of the codewords generated at each sensor in input reconstruction.

\subsection{Distributed Inference Accuracy}
\label{subsec:res_dist_detec}

\subsubsection{Comparative Evaluation}

\begin{table}[t]
\caption{The classification accuracy of the proposed framework under different compression ratios compared with different work from the literature on WARD dataset.}
\resizebox{\columnwidth}{!}{
\begin{tabular}{l|l|l|l}
\hline \hline
\multicolumn{1}{c|}{\textbf{Method}} & \multicolumn{1}{c|}{\textbf{CR=2}} & \multicolumn{1}{c|}{\textbf{CR=4}} & \multicolumn{1}{c}{\textbf{CR=8}} \\ \hline \hline
Cheng et al. (ASRCM) \cite{cheng2017accelerated}            & 94\%                                 & 88\%                                 & 83\%                                 \\ \hline
Cheng et al. (NN) \cite{cheng2017accelerated}                    & 82\%                                 & 78\%                                 & 75\%                                 \\ \hline
Zhang et al. \cite{zhang2013human}                          & 87\%                                 & 83\%                                 & 80\%                                 \\ \hline
Our Framework                         & \textbf{99.7\%}                                 & \textbf{97.4\%}                               & \textbf{95.6\%}                               \\ \hline
\end{tabular}
}
\label{table:accur_cr}
\end{table}

\begin{table}[t]
\caption{The classification accuracy of the proposed framework at compression ratio (CR=2) compared with different work from the literature on WARD dataset.}
\resizebox{\columnwidth}{!}{
\begin{tabular}{l|l}
\hline \hline
\textbf{Method}       & \textbf{Detection Accuracy} \\ \hline \hline
Zhu et al. \cite{zhu2019cost}                      & 99.00\%                        \\ \hline
Yang et al. \cite{yang2009distributed}        & 93.60\%                        \\ \hline
Huynh \cite{huynh2008human}                   & 96.97\%                      \\ \hline
He et al. + PCA \cite{he2012recognition}      & 76.31\%                      \\ \hline
He et al. + LDA \cite{he2012recognition}      & 40.30\%                    \\ \hline
He et al. + GDA \cite{he2012recognition}      & 99.20\%                    \\ \hline
Guo (Majority voting) \cite{guo2012human}     & 94.96\%                       \\ \hline
Guo (Maximum) \cite{guo2012human}             & 96.20\%                       \\ \hline
Guo (WLOP) \cite{guo2012human}                & 98.02\%                       \\ \hline
Guo (WLOGP) \cite{guo2012human}               & 98.78\%                       \\ \hline
Sheng et al. \cite{sheng2016short}            & 95.90\%                          \\ \hline
Oniga and Jozef \cite{oniga2015optimal}                               & 98.10 \%                        \\ \hline
\textbf{Our Framework}         & \textbf{99.7}\%                        \\ \hline \hline
\end{tabular}
}
\label{table:WARD_accuracy}
\end{table}

To evaluate the effectiveness of the proposed framework, we used a publicly available dataset called \textit{Wearable Action Recognition Database (WARD)} presented in \cite{yang2009distributed}. The obtained performance is compared against three other literature works applied to the same dataset. The dataset is designed for human activity recognition from sensors data. This dataset is collected from five sensor-boards attached to different points in the human body. Each sensor board has a tri-axial accelerometer and a bio-axial gyroscope with three and two dimensional outputs respectively. Each human operator performs 13 different actions which represent the labels (classes) predicted by the classifier at the FC.

\begin{figure}[t]
\centerline{\includegraphics[width=0.45\textwidth]{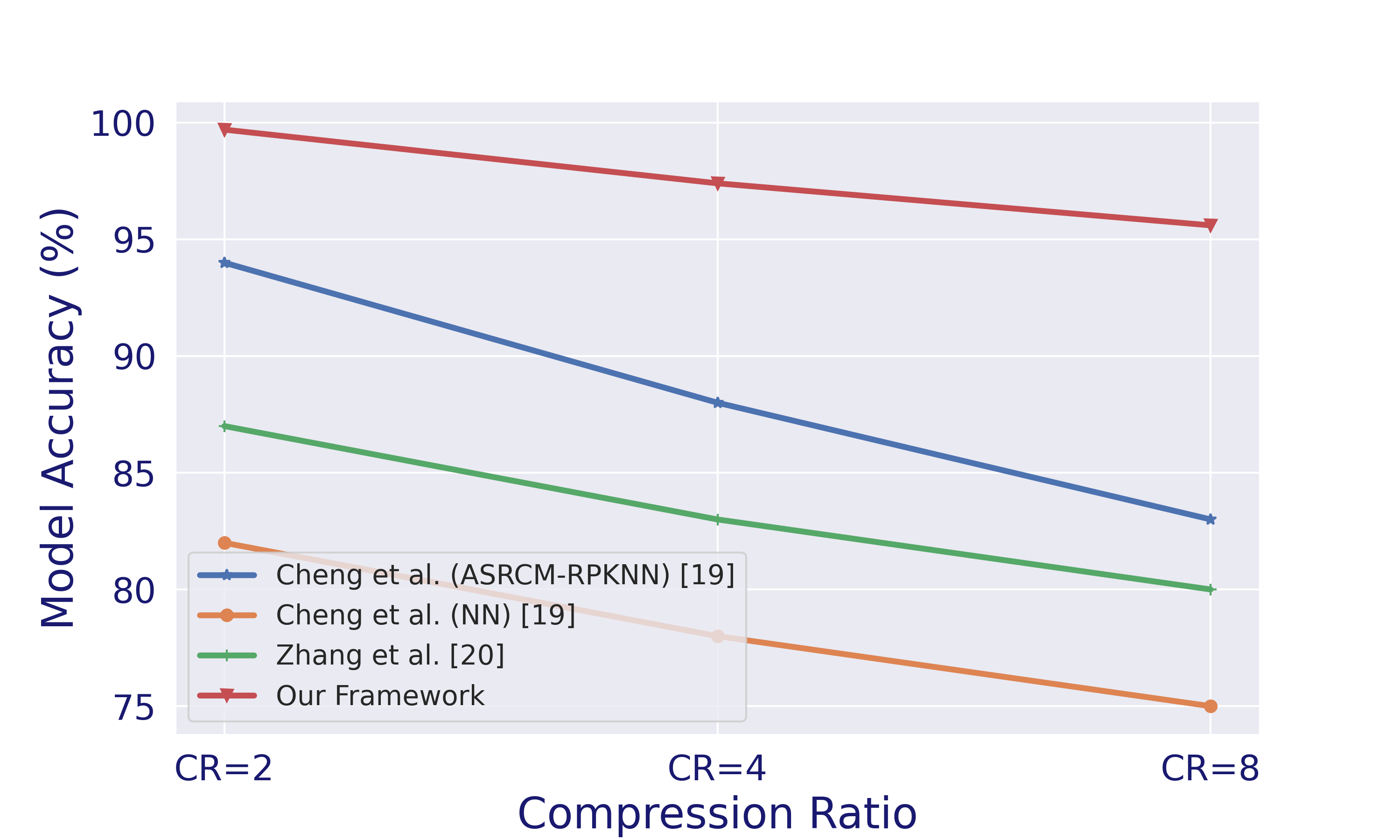}}
\caption{Comparison of model accuracy under different compression ratios.}
\label{fig:cr_accuracy}
\end{figure}

Table \ref{table:accur_cr} and Fig. \ref{fig:cr_accuracy} show a comparison between the performance of the proposed framework and other works in literature under different compression ratios. We can see from the table that the performance of our framework outperforms other works under all compression ratios. We can see that our framework preserves high accuracy even under high compression ratios. For example, increasing the compression ratio from 2 to 8 decreased the accuracy only 4.1\% from 99.7\% to 95.6\%. This is a small margin compared with 11\% loss in Cheng et al. (ASRCM) \cite{cheng2017accelerated}, and 7\% in Cheng et al. (NN) \cite{cheng2017accelerated} and Zhang et al. \cite{zhang2013human}.

Table \ref{table:WARD_accuracy} shows the classification accuracy of the proposed framework compared with the accuracy of other literature works. The table reports results for Zhu et al. \cite{zhu2019cost}, Yang et al. \cite{yang2009distributed}, Huynh \cite{huynh2008human}, He et al. \cite{he2012recognition}, Guo et al. \cite{guo2012human}, Oniga et al. \cite{oniga2015optimal}, and Sheng et al. \cite{sheng2016short}.
It is clear from the table that the proposed framework achieves state-of-the-art accuracy compared with other work in the literature. In addition, the proposed framework involves the minimal required bit rate, $R$, from the sensor nodes to the FC, which highly contributes to power saving and prolonging sensors lifetime. These results can be attributed to the fact that we learn complementary features between correlated sensors that highly contribute to improving the decision accuracy rather than learning local features for each sensor. This learning behaviour is motivated by the proposed loss function. Moreover, our framework jointly learns a quantizer function $q: \chi \rightarrow Z$ with the encoder function which minimizes the end-to-end error and improves the sensing task accuracy. Note that the work in \cite{zhu2019cost} explores the correlation between the sensor observations to disable the transmission on the sensors that did not capture new relevant features and thus save the consumed bandwidth. Comparatively, in our work, we exploit this correlation to transmit only the relevant complementary features. Consequently, we contribute in two directions, namely, saving the consumed bandwidth and, at the same time, improving the decision accuracy.

\subsubsection{Artificial Problem}

\begin{figure}[t]
\centerline{\includegraphics[width=0.50\textwidth]{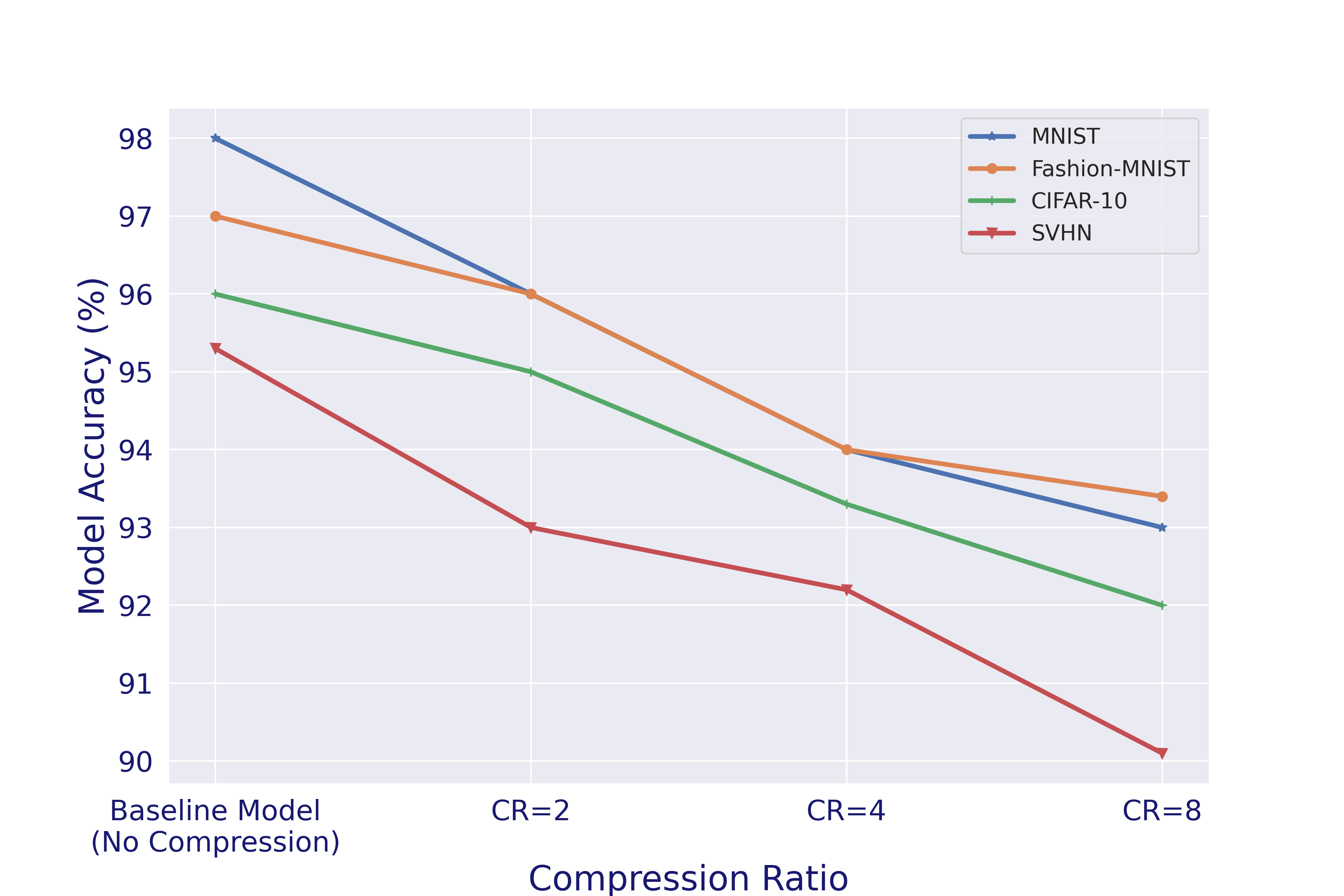}}
\caption{The decision accuracy of the proposed framework with four datasets under three different compression ratios. The baseline model represents the case in which we fuse the raw-observations to the FC without compression.}
\label{fig:acc_barchart}
\end{figure}

We tested the proposed framework with four datasets, which are: 1) MNIST \cite{MNIST}, 2) Fashion-MNIST \cite{xiao2017fashion}, 3) Street View Houses (SVH) \cite{netzer2011reading}, 4) CIFAR-10 \cite{krizhevsky2009learning}. These datasets represent different environments and generating distributions. For each dataset, we used different Compression Ratios, $CR$. $CR$ is defined as the ratio between the uncompressed dimension and compressed dimension \cite{sayood2017introduction}. It is worth to note that the compression ratios of the literature work consider only compression by dimensionality reduction (i.e., any input or output dimension $\in \mathbb{R}$). Based on that, the input and output space remains the same. Unlike prior methods, we go beyond to counts for the quantization (since an input dimension is $\in \mathbb{R}$ while an output dimension is quantized $\in \{0,1\}$). In these experiments, we simulate two sensors ($s_1$,$s_2$) sending their data to a FC. Assume the observations at sensor $s_1$ belongs to a class $C_i$ and at sensor $s_2$ belongs to a class $C_j$. The decision rule at the FC can be defined as:
\begin{equation}
\psi ( z^1, z^2) = \left \{
\begin{matrix}
 i &  if i=j\\ 
 -1 & if i\neq j  
\end{matrix} \right.
\end{equation}

\noindent In other words, the decision will be the class label if the two observations belong to the same label, and -1 otherwise. Since each dataset consists of images belonging to one out of 10 total classes, we expect the classifier to have 11 classes.

In order to make fair comparison between the framework accuracies with different compression ratios, we used the same classifier capacity (in terms of number of layers, the nodes in each layer, the activation functions used, etc.) for each dataset. We compared the obtained results with the baseline model accuracy. The baseline model is defined as the accuracy of a neural network classifier taking as input the raw observations without compression or quantization, $x^t$. In this case, the FC has the complete vector of sensed data, which represents the optimal case in terms of the data availability at the FC. 

Fig. \ref{fig:acc_barchart} shows the obtained results in each case. We can notice that the framework performance approaches the baseline with the lowest compression ratio, $CR=2$ in the table. A small loss in the accuracy is reported with higher $CR$ (i.e., $CR=4$ and 8). However, the obtained accuracy is still high even with the highest $CR$. For example, we obtained 95.3\% of the baseline with $CR=8$ in MNIST dataset. Which means compressing the observations to just 12.5\% of its original dimension with quantization, results in 4.7\% reduction in accuracy.

\begin{figure}[t]
\centerline{\includegraphics[width=0.45\textwidth]{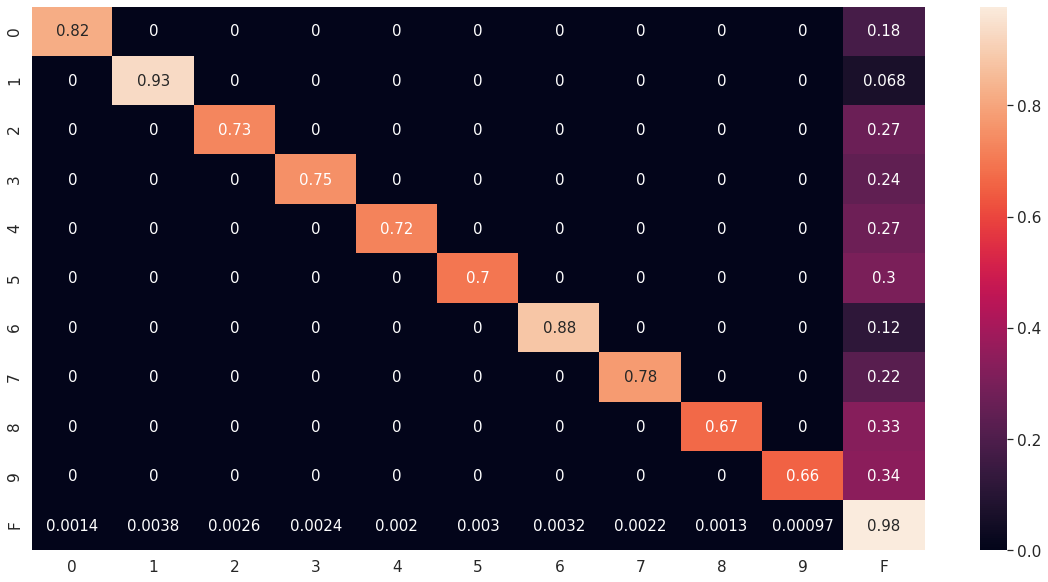}}
\caption{A heatmap representation for the confusion matrix of an MNIST classifier with 98-dimension latent code corresponding to compression ratio of eight. The class label F here represent class label -1.}
\label{fig:conf_mat_heatmap}
\end{figure}

In the reconstruction of the training dataset, we randomly shuffle the datasets in each sensor. Consequently, most of the observations combinations fall in the class of -1 which means the two observations are not in the same class. This produces an imbalanced class distributions. Due to this imbalance, we report the confusion matrix of the framework classifier for MNIST dataset and 98-dimension latent code. Fig. \ref{fig:conf_mat_heatmap} shows that the model is capable of taking the right decision with high accuracy even with imbalanced data. 

The key idea of compressing correlated sensors data is extracting complementary information from correlated observations and ignoring any redundancies. Our proposed loss function (Eq \ref{eq:proposed_loss}) achieved this goal by incorporating a KL divergence term to the loss function between the soft labels generated by a baseline model (e.g., a large model trained on raw observations to predict $P(Y|X)$) and the decision function at the FC ($P(Y|Z)$). To minimize this term, we encode only the complementary features from each sensor which the help of the FC that behaves as the baseline model. As described in Algorithm. \ref{algo:training}, we jointly train the encoder models at each sensor with the decision function at the FC in an end-to-end fashion. This end-to-end training makes the encoders jointly learn these features with the decision function as they receive penalization based on the distance between the predicted distribution and that of the baseline model.

\begin{figure}[t]
\centerline{\includegraphics[width=0.5\textwidth]{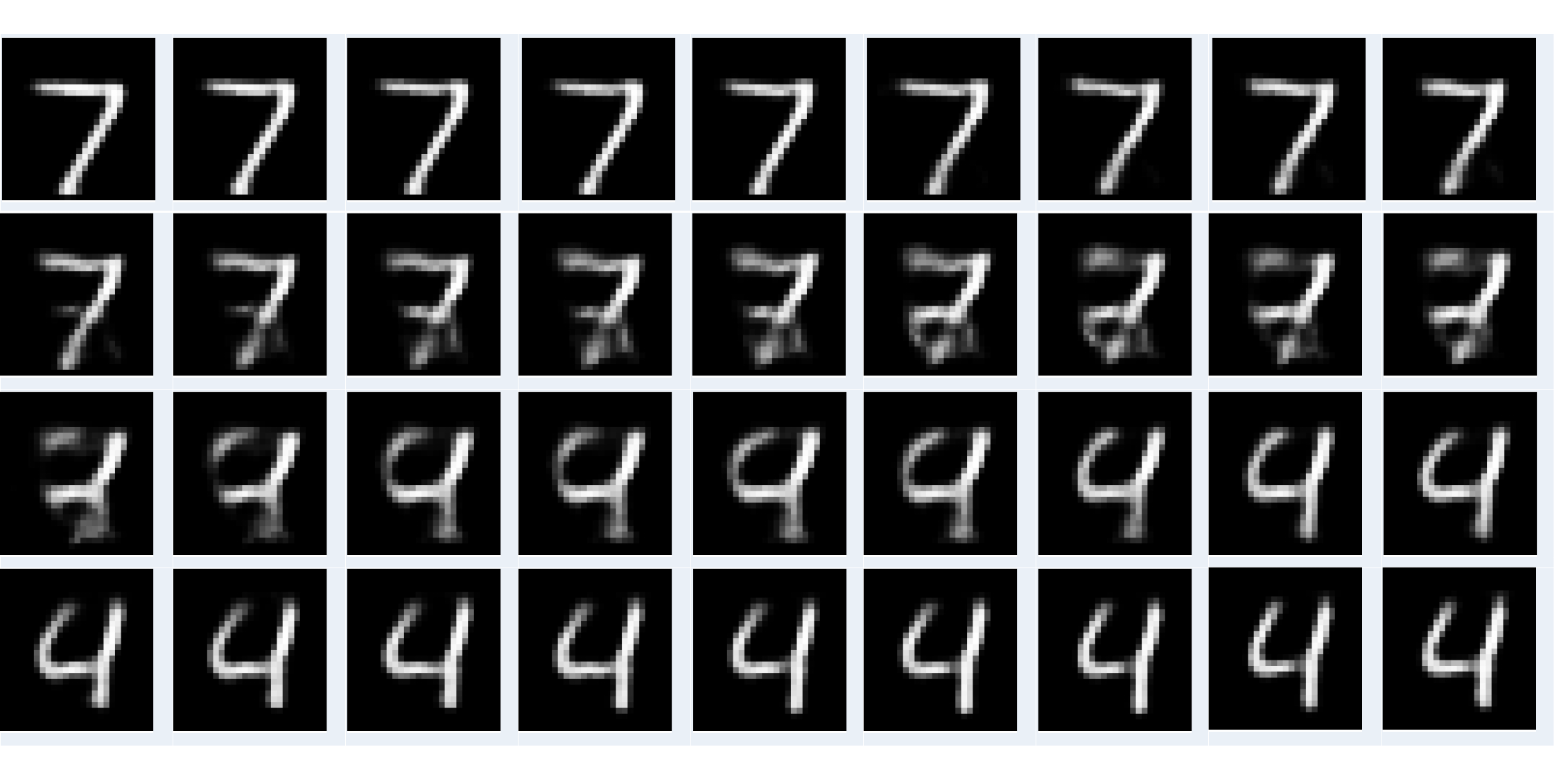}}
\caption{Interpolation between two points in the latent space. We choose a start and end points, then we gradually flip a bit each time along the different bits between the two vectors. The starting point shown in the top-left corner, the end point in the bottom-right.}
\label{fig:interpolation}
\end{figure}

\subsection{Semantics of the Latent Representation}
\label{subsec:res_sem_rep}

\begin{figure*}[t]
\centerline{\includegraphics[width=0.9\textwidth]{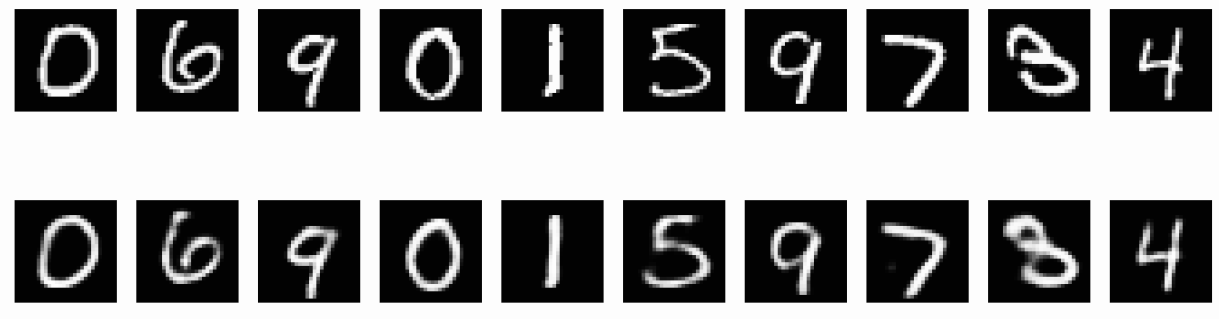}}
\caption{Reconstruction of MNIST data images using a 98-dimensional discrete latent code. The top row shows the original input, while the reconstruction is shown in the bottom row.}
\label{fig:res_reconstruction}
\end{figure*}

In autoencoders-based architectures for dimensionality reduction, a special interest is paid to the robustness of the learned codewords in the latent space \cite{choi2019neural}. To evaluate the robustness of such codewords, we interpolate between different points in the latent space and observe, qualitatively, the gradual changes in the reconstructed data. This widely used experiment verifies that the model: (a) has injected enough redundancies into the codewords and consequently the model is capable for reconstructing the input even in the presence of errors in the codeword, (b) has learned relevant features of the underlying structure of the data.

We randomly select two test points to represent the start and end points. Each step, we flip a bit in the latent codeword, fed the new obtained codeword to the decoder model and observe the gradual changes in the reconstruction. Algorithm \ref{algo:semantics}, describes this experiment in more details. Fig. \ref{fig:interpolation} shows the gradual transition in the digit shape with the gradual bit flipping. We can observe that decrementing the hamming distance between the start and end points, each bit-flip, slowly alters the characteristic features of digit until the digit reaches the end point.

\subsection{Rate/Computation Tradeoff}
\label{subsec:trandeoff}

\begin{figure}[t]
\centerline{\includegraphics[width=0.5\textwidth]{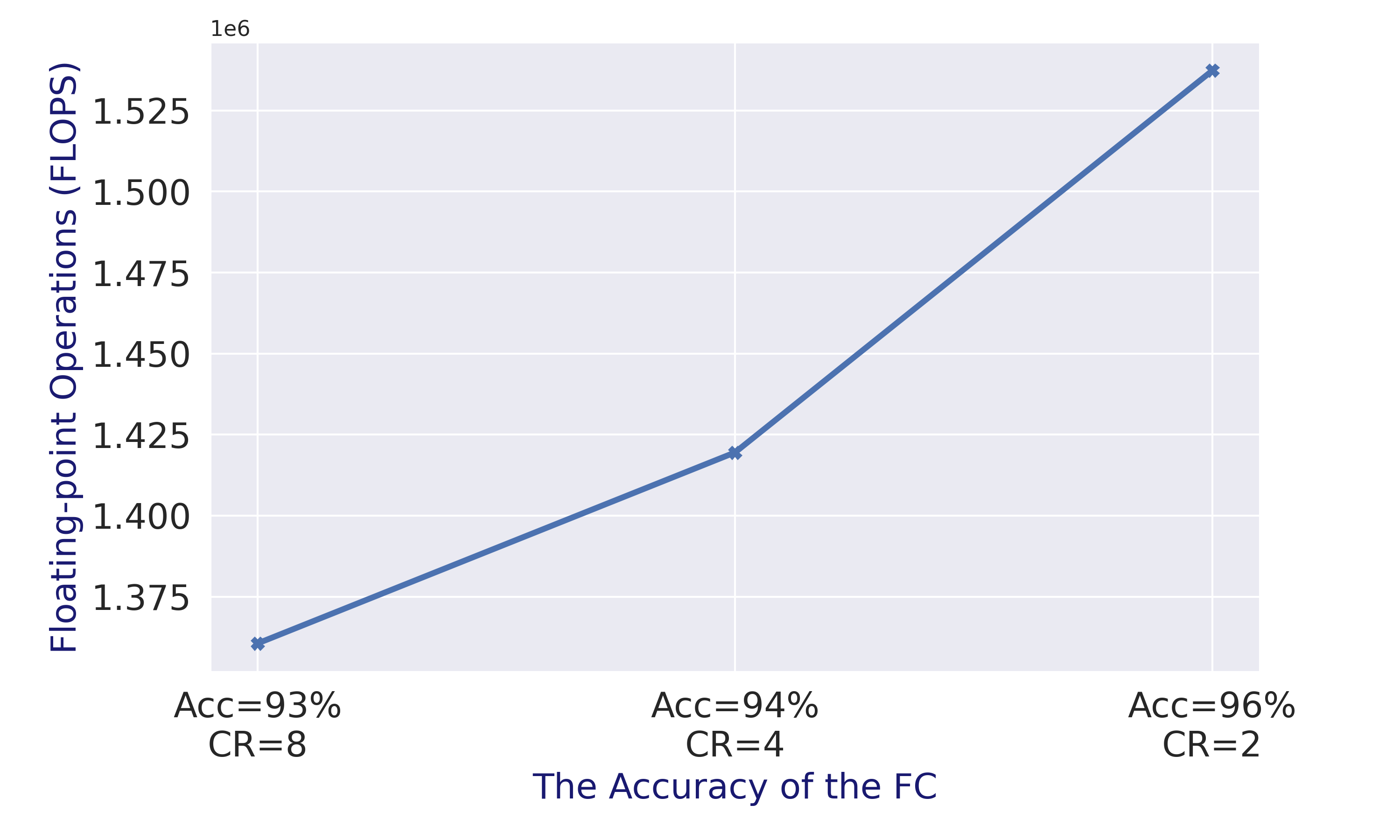}}
\caption{Tradeoff between model accuracy (achieved at different compression ratios) and the consequent increase of computation requirement.}
\label{fig:FLOPS_tradeoff}
\end{figure}

Although observation compression reduces the consumed bandwidth for transmission, this reduction comes with the cost of computation and accuracy. The required computation resources (measured by the floating-point operations (FLOPS)) increases according to the model complexity (measured by the number of weights). Moreover, increasing the model complexity leads to improved compression, and consequently improved decision accuracy at the FC. Therefore, a design decision should compromise between the model complexity on the one hand, and the consumed computations and FC accuracy on the other hand. However, the training phase can be done off-line (before the deployment of the sensors), and only the inference will take place during the operation which requires only one forward pass (a very small number of FLOPS) to predict the encoded messages. Fig. 9 shows this trade-off trend between the computation requirement (measured by FLOPS) and the model accuracy. In this figure, we can see that increasing the model accuracy requires adopting smaller compression ratios which implies higher data transmission. On the other hand, a smaller compression ratio requires transmitting more data and requires more computational resources at each sensor. The optimization of compression ratios is out of the scope of this paper, and will be explored in our future work.

\vspace{-0.08cm}
\subsection{Applicability}
\label{subsec:Applicability}

\begin{figure}[t]
\centerline{\includegraphics[width=0.5\textwidth]{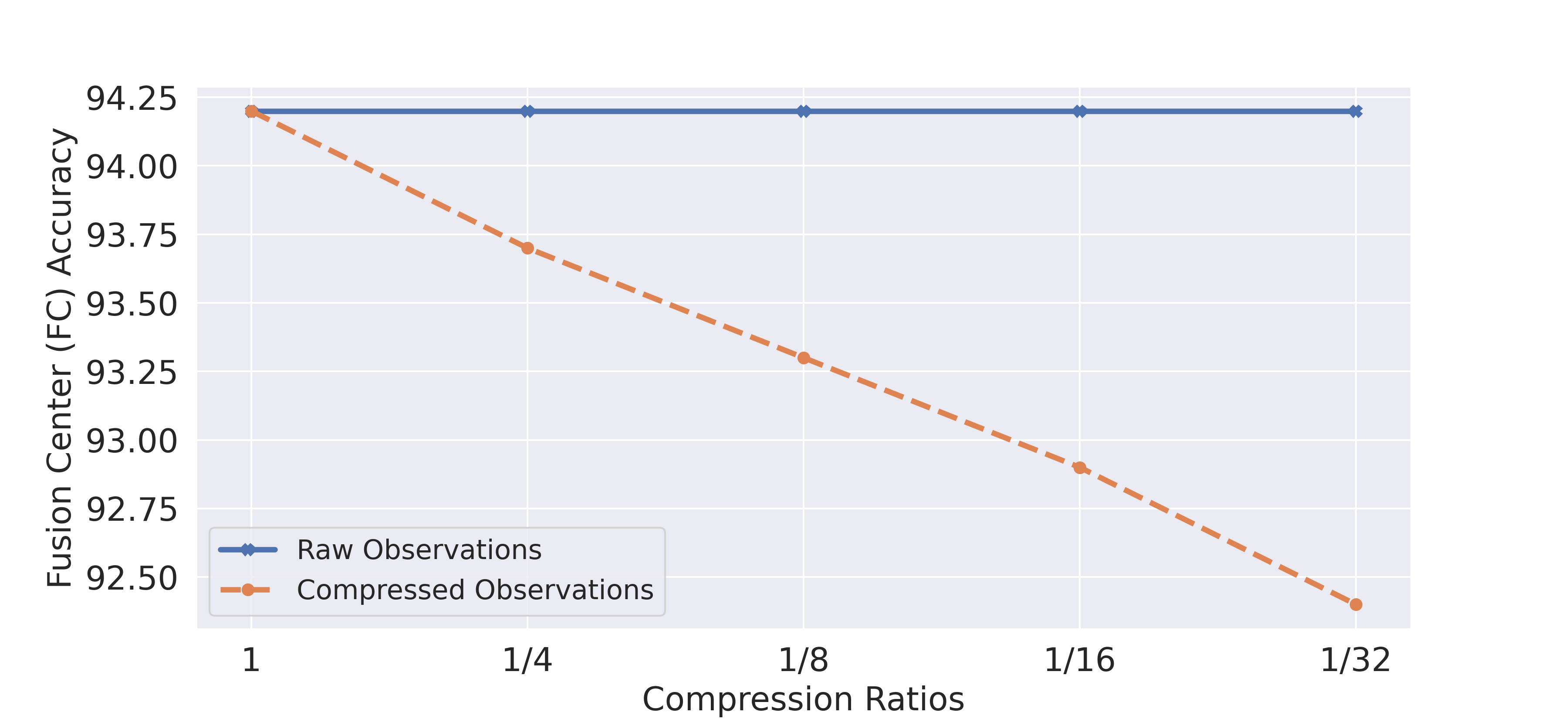}}
\caption{The decision accuracy for a wireless link adaptation problem under different compression ratios. The results confirm the general applicability of the proposed system to problem from different domains.}
\label{fig:aaplicability}
\end{figure}

Our proposed framework along with the proposed loss function and training procedure shown in Algorithm. 1 can work with any type of parallel distributed detection network. This type of settings has various applications in wireless IoT. Although minor customizations are required to fit each specific problem, the framework still widely applicable to various problems from various domains. In this paper, we reported the experimental results on various types of sensors and applications (e.g., image classification, human activity recognition, etc.). Specifically, we experimented 5 different datasets (MNSIT, Fashion MNIST, SVHN, CIFAR-10, WARD) representing three different types of sensors (cameras, gyroscope, and accelerometer). To verify the generality of our framework, we tested the proposed framework in a completely different domain, i.e., wireless link adaptation using three datasets combined in a global dataset \cite{hussien2021towards}. In this scenario, the sensors are the antennas at each mobile node, the observations are the channel state information (CSI) captured at each mobile, and the environment that the sensors monitoring is the wireless channel \cite{hussien2020prvnet}. The sensors send their observations to a FC to take a global decision. The base station (BS) acts as a FC in this case, and the decision is the selected modulation and coding scheme. The results shown in Fig. \ref{fig:aaplicability} show minor loss in the adaptation decision at the FC with the increase in the adopted compression ratio. For example, when compressing the original raw observations (i.e., CSI in this case), the accuracy only drops from 94.25\% to 93.7\%. This means only 0.55\% loss in accuracy is achieved while saving 75\% of the original bandwidth. The obtained results confirm the general applicability of our proposed method in different domains and problems. 

\subsection{Results of Input Reconstruction}
\label{subsec:res_decomp}
To further evaluate the robustness of the learned features, we evaluated the proposed model in input reconstruction task. In this experiment, MNIST and Fashion-MNIST datasets are used for evaluation. We used a $CR$ equals to eight which corresponds to a latent code of 98-bit. Fig. \ref{fig:res_reconstruction} shows the result of the input reconstruction. 

\section{Related Work}
\label{sec:related_work}
In this section, we review prior work that is related to various
aspects of this paper.

A similar work has been proposed for specific problems. For example, a line of work has been proposed for the problem of human activity recognition \cite{yang2008distributed, guo2012human, huynh2008human, he2012recognition}. In this problem, the hypotheses are the different human actions, while the data comes from multiple sensors fixed on the actor body (e.g., gyroscope, accelerometer, etc.). This work focuses on compressing the sensors data without hurting the recognition accuracy. For example, in \cite{ yang2008distributed }, the authors aimed to achieve a high action-classification accuracy with the minimum bandwidth consumption. At each sensor, the decision is taken based on its local information. The FC then takes a global decision using a majority-voting mechanism. Although they obtained a good results, this approach ignores any complementary information captured by other sensors. Another work has been proposed for the problem of earthquakes detection from wireless IoT sensors network \cite{faulkner2011next}. They presented a distributed approach for rapid detection of earthquakes using cell phone accelerometers, consumer USB devices, and cloud computing based sensor fusion. The approach proposed in \cite{faulkner2011next} learns a threshold for each sensor involved in the network in a way that maximizes the performance of the anomaly detection algorithm employed at the FC. Experimental results showed that the proposed approach successfully distinguished between seismic motion from accelerations due to normal daily manipulation.

The work in \cite{raghavan2019binary} studied the problem of binary hypothesis testing with two observers, where the collected observations are assumed to be statistically correlated. To reach a decision, one of three solutions can be adopted. The first is a centralized solution in which the observations collected by both observers are sent to the FC. A global decision is taken at the FC based on the received sensor observations. The main concern of this solution is the huge bandwidth incurred in observation transmission. The second solution makes each observer rely on its own locally collected observation. Then, each node exchanges its locally obtained decision with other sensors to reach a global decision. The main limitation of this solution is that each sensor depends only on its local information and ignores any complementary information captured by other sensors. In the last solution, each observer formulates the problem as a sequential hypothesis-testing problem. The authors in \cite{bouchoucha2015distributed} proposed a framework for exploiting the correlation between observations to reduce the mean square error of the distributed estimation. Specifically, each node predicts its next observation and transmits the quantized prediction errors (innovations) to the FC instead of the quantized observations.

In the context of task-aware compression, a similar problem has been addressed in \cite{chinchali2018neural, hu2020starfish, amer2020image}. For example, the authors in \cite{chinchali2018neural} used a reinforcement agent at each sensor node to compress the observations before feeding them to the FC. The reward function at each agent considers its commitment to the assigned bandwidth. Although they achieved a good performance, there is a probability that the agent does not meet the bandwidth constraints after deployment. While in \cite{hu2020starfish}, the authors proposed \textit{Starafish}, an image compression framework that outperforms JPEG by by up to 3X in terms of bandwidth consumption and up to 2.5X in power consumption. The authors in \cite{hu2020starfish} used an AutoML technique to search for tiny ML models that can work on power AIoT accelerators.

We can summarize the limitations of the literature work, which we addressed in our work, as: 1) the conditional-independence assumption of the sensor observations does not usually hold; 2) the conditional-independence assumption ignores the potential opportunity to benefit from complementary features captured by different sensors; 3) the compression algorithms are designed independently from the sensing goal; 4) the limited power of analytical-based techniques in dealing with large number of possible decisions and correlated sensors.


\section{Conclusion}
\label{sec:conslusion}
In this paper, we proposed a deep learning framework for compressing correlated sensor observations in distributed inference problems. The proposed framework employ discrete representation autoencoders to encode the observations at each sensor. A novel loss function is proposed to improve the accuracy of the framework. A multi layer perceptron architecture has been used at the FC to jointly-learn the decision rule. The proposed framework addresses the hard to tackle problem of correlated sensor observations and does not assume any prior knowledge about the  distribution of these observations. The performance of the model has been extensively verified using different datasets and proved to provide significant performance improvement.


\bibliographystyle{./bibliography/IEEEtran}
\bibliography{./bibliography/IEEEabrv,./bibliography/References}

\end{document}